\documentclass[a4paper,11pt]{article}
\usepackage[a4paper, total={17cm, 25cm}]{geometry}
\usepackage{fancyhdr}
\usepackage{graphicx}
\usepackage{subfigure}
\usepackage[square,sort,comma,numbers]{natbib}
\fancypagestyle{plain}{
\fancyhf{} 
\fancyhead[RH]{\thepage}
\fancyfoot[CF]{28th Texas Symposium on Relativistic Astrophysics \\ Geneva, Switzerland -- December 13-18, 2015}

}
\def \f{\frac}
\def \w{\wedge}
\def \o{\omega}

\def \a{\alpha}

\def \t{\tilde}
\def \O{\Omega}
\def \g{\gamma}
\def \s{\sigma}
\def \e{\epsilon}
\def \p{\partial}
\def \d{\Delta}

\def \a{\alpha}

\def \th{\theta}
\def \r{\rho}

\begin{document}
%%%%%%%%%%%%%%%%%%%%%%%%%%%%%%%%%%%%%%%%%%%%%%%%%%%%%%%%%%%%%%%%%%%%%%%%%%%%
%%%% Title

\title{\bf Frame-dragging Effect in Strong Gravity Regime}

\author{
Chandrachur Chakraborty \\ Tata Institute of Fundamental Research, Mumbai 400005, India
}

\date{}

\maketitle

\begin{abstract}
The exact frame-dragging (or Lense-Thirring (LT) precession) rates for Kerr, Kerr-Taub-NUT (KTN) and Taub-NUT 
spacetimes have been derived. Remarkably, in the case of the `zero angular 
momentum' Taub-NUT spacetime, the frame-dragging effect is shown not to vanish, 
when considered for spinning test gyroscope. In the case of the interior of the 
pulsars, the exact frame-dragging rate monotonically decreases 
from the center to the surface along the pole and but it shows an `anomaly' along the equator. 
Moving from the equator to the pole, it is observed that this `anomaly'
disappears after crossing a critical angle. The `same' anomaly can also be found 
in the KTN spacetime. The resemblance of the anomalous LT precessions 
in the KTN spacetimes and the spacetime of the pulsars could be used to identify a role of Taub-NUT solutions
in the astrophysical observations or equivalently, a signature of the existence of NUT
charge in the pulsars.
\end{abstract}
%%%%%%%%%%%%%%%%%%%%%%%%%%%%%%%%%%%%%%%%%%%%%%%%%%%%%%%%%%%%%%%%%%%%%%%%%%%%

%%%% Paper body

\section{Introduction}

Stationary spacetimes with angular momentum (rotation) are known 
to exhibit an effect called Lense-Thirring (LT) precession whereby 
locally inertial frames are dragged
along the rotating spacetime, making any test gyroscope in such spacetimes
{\it precess} with a certain frequency called the LT precession
frequency \cite{lt}. This frequency has been shown to decay as the 
inverse cube of the
distance of the test gyroscope from the source for large 
enough distances where curvature effects are small,
and known to be proportional to the angular momentum of the
source. The largest precession frequencies are thus expected to be
seen very close to the source (like the surface of a
pulsar, or the horizon of a black hole), as well as for spacetimes
rotating very fast with large angular momenta. 

Earlier analyses of the LT effect \cite{jh} assume slowly rotating
($r>>a,~ a$ is the Kerr parameter of the rotating spacetime due to a compact
object like a black hole) spacetime for the test gyroscope \cite{schiff}. Thus,
the rotating spacetime solution 
is usually approximated as a Schwarzschild spacetime, and the effect
of rotation is confined to a perturbative term added on to the Schwarzschild
metric. This leads to the standard result for LT precession
frequency in the weak field approximation, given by \cite{jh}
\begin{equation}
\vec{\Omega}_{LT}=\f{1}{r^3}[3(\vec{J}.\hat{r})\hat{r}-\vec{J}]
\label{we1}
\end{equation}
where, $\hat{r}$ is the unit vector along $r$ direction.
In a recent work reported in ref. \cite{ks}, an alternative approach based
on solving the geodesic equations of the test gyroscope numerically,
{\it once again} within the weak gravitational field approximation, is used to compute the
frame-dragging effect for galactic-centre black holes.

In another very recent related work \cite{hl}, Hackman and Lammerzahl have given an
expression of LT precession (orbital plane precession) valid up to {\it first order} in the Kerr
parameter $a$ for a general axially symmetric Plebanski-Demianski
spacetime. This is obviously a good approximation for slowly-rotating
compact objects. The LT precession rate has also been derived \cite{gk} through solving 
the geodesic equations for both Kerr and Kerr-de-Sitter spacetimes 
at the {\it polar orbit} but these results are not applicable for orbits
which lie in orbital planes other than the polar plane. We understand that 
observations of precession due to locally inertial frame-dragging 
have so far focused on spacetimes where the curvatures
are small enough; e.g., the LT precession in the earth's gravitational
field which was probed recently by Gravity Probe B \cite{grav}. There has been so far no
attempt to measure LT precession effects due to frame-dragging in
strong gravity regimes.
 
 Two motivating factors may be cited in 
support of such a contention.
First of all, the near-horizon physics of black holes and that of 
the outer layers of neutron stars emitting X-rays from their accretion 
discs also might need to be reanalyzed in view of the nontrivial LT 
precession of test geodesics in their vicinity. With upcoming X-ray observatories,
as well as multi-wavelength strong gravity space probes currently under
construction, which envisage to make observations of possible frame-dragging 
effects in strong gravity situations in the near future,
the need to go beyond the weak field approximation is paramount. 
A recent work by Stone and Loeb \cite{sl} has estimated the effect of 
weak-field LT precession on accreting matter close to compact accreting 
objects. While there are claims that what has been estimated in this 
work pertains more to orbital plane precession, rather than precession of a 
test gyroscope (which remains the classic example of LT precession), 
it is obvious that in the vicinity of the spacetime near the surface of 
pulsars (respectively, the horizons of black holes), the large LT 
precession of test gyroscopes ought to manifest in changes in the predicted
X-ray emission behaviour originating from modifications in the behaviour 
of infalling timelike geodesics of accreting matter particles due to the LT 
precession. Thus, there is sufficient theoretical motivation to compute 
LT precession rates in the
strong gravity regime, in a bid towards a prediction that future probes
of the inertial frame dragging effect, in such a regime, may correlate with.

\section{Exact LT precession frequency in stationary spacetime \& its applications}
The exact LT precession frequency of a test gyroscope in strongly curved stationary 
spacetimes, analyzed within a `Copernican' frame, is expressed 
as a co-vector given in terms
of the timelike Killing vector fields $K$ of the stationary spacetime,
as (in the notation of ref. \cite{ns})
\begin{eqnarray}
\t \O&=&\f{1}{2K^2}*(\t K \w d\t K) 
~\label{ltfr}
\end{eqnarray} 
where, $\t K$ \& $\t \O$ denote the one-form dual to $K$ \& $\O$, respectively. 
Note that $\t \O$ vanishes if and only if $(\t K \w d\t K)=0$. This happens 
only for a static spacetime.

Using the coordinate basis form of $ K= \p_0$,
the co-vector components are easily seen to be $K_{\mu} =
g_{\mu 0}$. Thus, the vector field corresponding to the LT precession co-vector
can be expressed in coordinate basis as   
\begin{eqnarray}
\O = \f{1}{2} \f{\e_{ijl}}{\sqrt {-g}} \left[g_{0i,j}\left(\p_l - 
\f{g_{0l}}{g_{00}}\p_0\right) -\f{g_{0i}}{g_{00}} g_{00,j}\p_l\right]
\label{s25}
\end{eqnarray}
The remarkable feature of the above equation (\ref{s25}) is that it is 
applicable to any arbitrary stationary spacetime (irrespective of
whether it is axisymmetric or not); it gives us the exact rate of LT
precession in such a spacetime. For instance, a `non-rotating' Newman-Unti-Tamburino \cite{nut}
(NUT) spacetime is known to be spherically
symmetric, but still has an angular momentum (dual or `magnetic' mass
\cite{rs}); we use  Eq.(\ref{s25}) to compute the LT precession
frequency in this case as well. This result is rather general, 
because, there is only one constraint on the spacetime : that it must 
be stationary, which is the only necessary condition for the LT precession. 
The utility of this equation is that; if any metric $(g_{\mu\nu})$ contains
all 10 $(4\times4)$ elements non-vanishing, it can be used to
calculate the LT precession in that spacetime.
 In this case, the precession rate depends only 
on non-zero $g_{0\mu} (\mu=0,1,2,3)$ components, not on any other non-zero off-diagonal 
components of the metric. Thus, this equation also reveals that the LT precession
rate is completely determined by the metric components $g_{0\mu}$,
and is quite independent of the other components (in co-ordinate basis).

Now, we should discuss that why a stationary spacetime shows this frame-dragging effect
irrespective of whether it is axisymmetric or not.
A spacetime is said to be stationary if it possesses a
timelike Killing vector field $\xi^a$; further, a stationary spacetime is said to
be static if there exists a spacelike 
hypersurface $\Sigma$ which is orthogonal to the orbits of
the timelike isometry. By Frobenius's theorem of 
hypersurface orthogonality, we can write for a static spacetime,
\begin{equation}
\xi_{[a}\nabla_b\xi_{c]}=0
\label{ho}
\end{equation}
If $\xi^a\neq 0$ everywhere on $\Sigma$, then in a neighbourhood
of $\Sigma$, every point will lie on a unique orbit of $\xi^a$
which passes through $\Sigma$. From the explicit form of a
static metric, it can be seen that the diffeomorphism defined
by $t\rightarrow -t$ (the map which takes each point on
each $\Sigma_t$ to the point with the same spatial
coordinates on $\Sigma_{-t}$), is an isometry. The ``time
translation'' symmetry, $t\rightarrow t+constant$ is 
possessed by all stationary spacetimes. Static 
spacetimes on the other hand, possess an additional symmetry, ``time 
reflection'' symmetry over and above the ``time
translation symmetry''. Physically, the fields which are
time translationally invariant can fail to be time reflection invariant
if any type of ``rotational motion'' is involved, since the time
reflection will change the direction of rotation. For example,
a rotating fluid ball may have a time-independent matter and
velocity distribution, but is unable to possess a time 
reflection symmetry \cite{rw}. In the case of stationary spacetimes,
the failure of the hypersurface orthogonality condition (Eq.\ref{ho})
implies that neighbouring orbits of $\xi^a$ ``twist'' around
each other. The twisting of the orbits of $\xi^a$ is the cause
of that {\it extra precession} in stationary non-static spacetimes.  

\subsection{Exact LT precession rate in Kerr spacetime}
One can now use Eq. (\ref{s25}) to calculate the precession rate of
a test gyroscope in a Kerr spacetime to get the LT precession in a strong gravitational
field. In Boyer-Lindquist coordinates, the Kerr metric is written as,
\begin{eqnarray}
ds^2=-\left(1-\f{2Mr}{\rho^2}\right)dt^2-\f{4Mar \sin^2\theta}{\rho^2}d\phi dt
+\f{\rho^2}{\Delta}dr^2  
+\rho^2 d\theta^2+\left( r^2+a^2+\f{2Mra^2 \sin^2\theta}{\rho^2}\right) \sin^2\theta d\phi^2 
\label{k1}
\end{eqnarray}
where, $a$ is Kerr parameter, defined 
as $a=\f{J}{M}$, the angular momentum per unit mass and
\begin{equation}
 \rho^2=r^2+a^2 \cos^2\theta,   \,\,\,\,     \\    \,\,\,\,      \Delta=r^2-2Mr+a^2 .
\label{k2}
\end{equation}
Now, using Eq.(\ref{s25}) we can obtain the following expression of 
LT precession rate in Kerr spacetime \cite{cm}
\begin{eqnarray}
\vec{\O}_{LT}^K= 2aM \cos\theta \f{r\sqrt{\Delta}}{\rho^3 (\rho^2-2Mr)}\hat{r} 
-aM \sin\theta \f{\rho^2-2r^2}{\rho^3(\rho^2-2Mr)}\hat{\th} .
\label{k10}
\end{eqnarray}
This is the LT precession rate where no weak gravity approximation has
been made. In slow-rotation limit $(r >> a)$, the Kerr metric 
is approximated as a Schwarzschild metric with 
the cross term $(g_{\phi t}d\phi dt)$, that is
\begin{equation}
 ds^2=ds_{Sch}^2-\f{4Ma \sin^2\theta}{r} d\phi dt
\label{w1}
\end{equation}
and Eq. (\ref{k10}) reduces to Eq. (\ref{we1}) which is quite well-known to us.

\subsection{Exact LT precession rate in Kerr-Taub-NUT spacetime}
The Kerr-Taub-NUT spacetime is geometrically a stationary, axisymmetric
vacuum solution of Einstein equation with Kerr parameter $(a)$ and
NUT charge $(n)$. If the NUT charge vanishes, the solution reduces to the
Kerr geometry. The metric of the Kerr-Taub-NUT spacetime is
\begin{equation}
ds^2=-\f{\d}{p^2}(dt-A d\phi)^2+\f{p^2}{\d}dr^2+p^2 d\th^2
+\f{1}{p^2}\sin^2\th(adt-Bd\phi)^2
\label{lnelmnt}
\end{equation}
with 
\begin{eqnarray}\nonumber
\d&=&r^2-2Mr+a^2-n^2,  p^2=r^2+(n-a\cos\th)^2,
\\
A&=&a \sin^2\th+2n\cos\th, B=r^2+a^2+n^2.
\end{eqnarray}
As the spacetime has an intrinsic angular momentum (due to the Kerr parameter $a$),
we can expect a non-zero frame-dragging effect which cam be written as
\begin{eqnarray}
\vec{\O}_{LT}^{KTN}=\f{\sqrt{\d}}{p}\left[\f{a \cos\th}{\r^2-2Mr-n^2}
-\f{a \cos\th-n}{p^2}\right]\hat{r}
+\f{a \sin\th}{p} \left[\f{r-M}{\r^2-2Mr-n^2}
-\f{r}{p^2}\right]\hat{\th}
\label{kt}
\end{eqnarray}
where, $\r^2=r^2+a^2\cos^2\th$. 
\begin{figure}[h]
  \begin{center}
\subfigure[along the pole]{
\includegraphics[width=1.9in,angle=0]{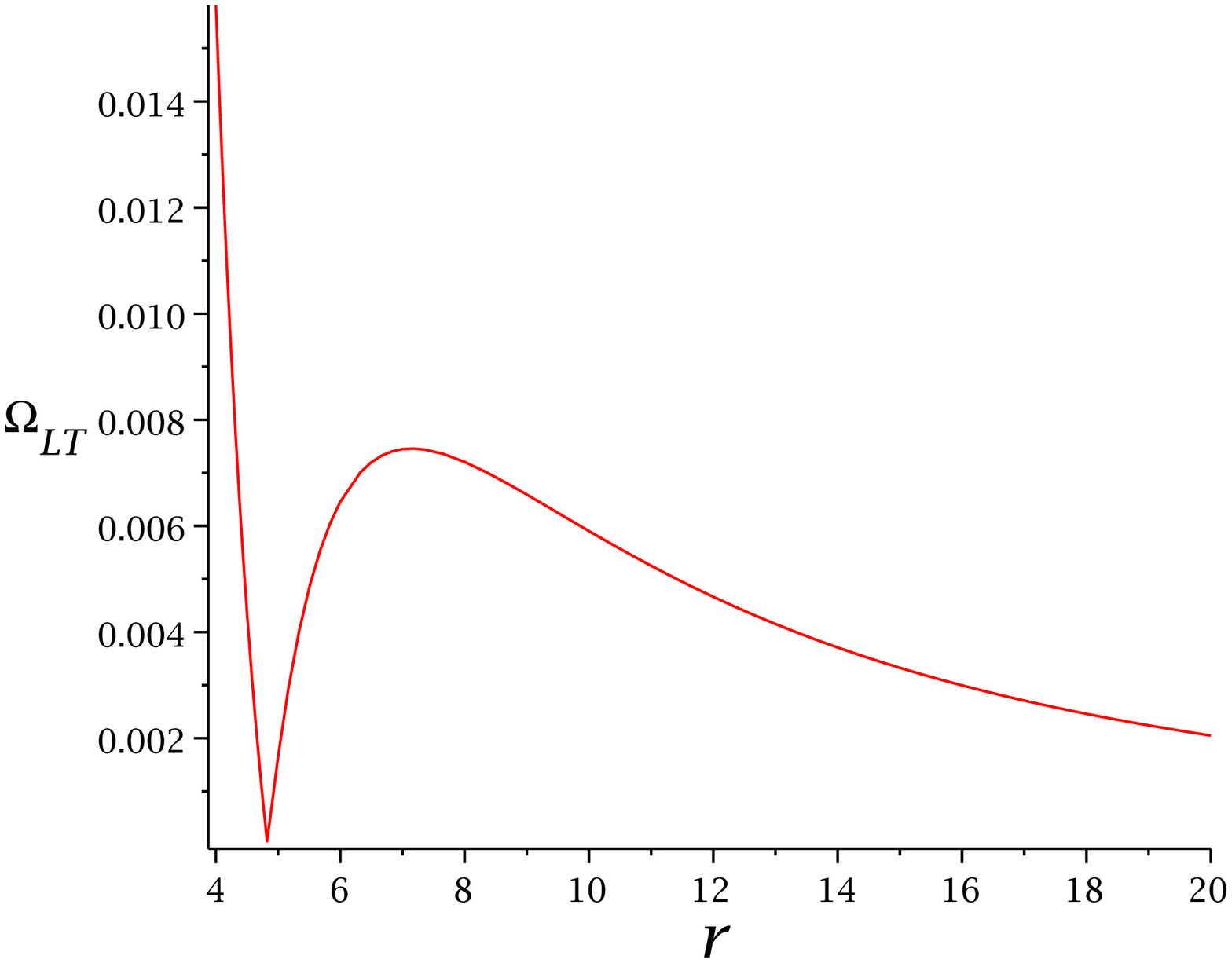}} 
\subfigure[along the equator]{
 \includegraphics[width=1.9in,angle=0]{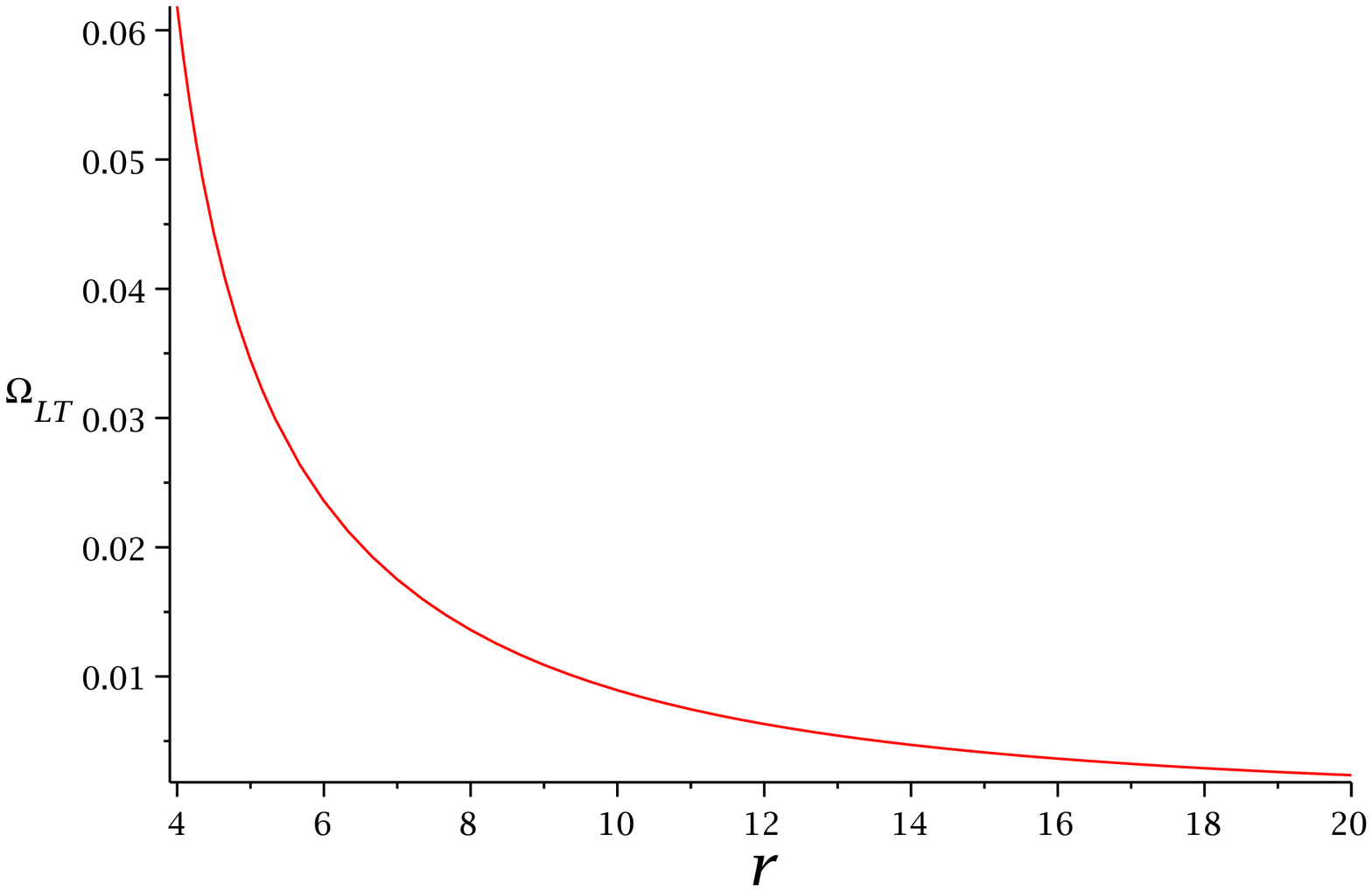}}
\subfigure[3-D]{
 \includegraphics[width=1.9in,angle=0]{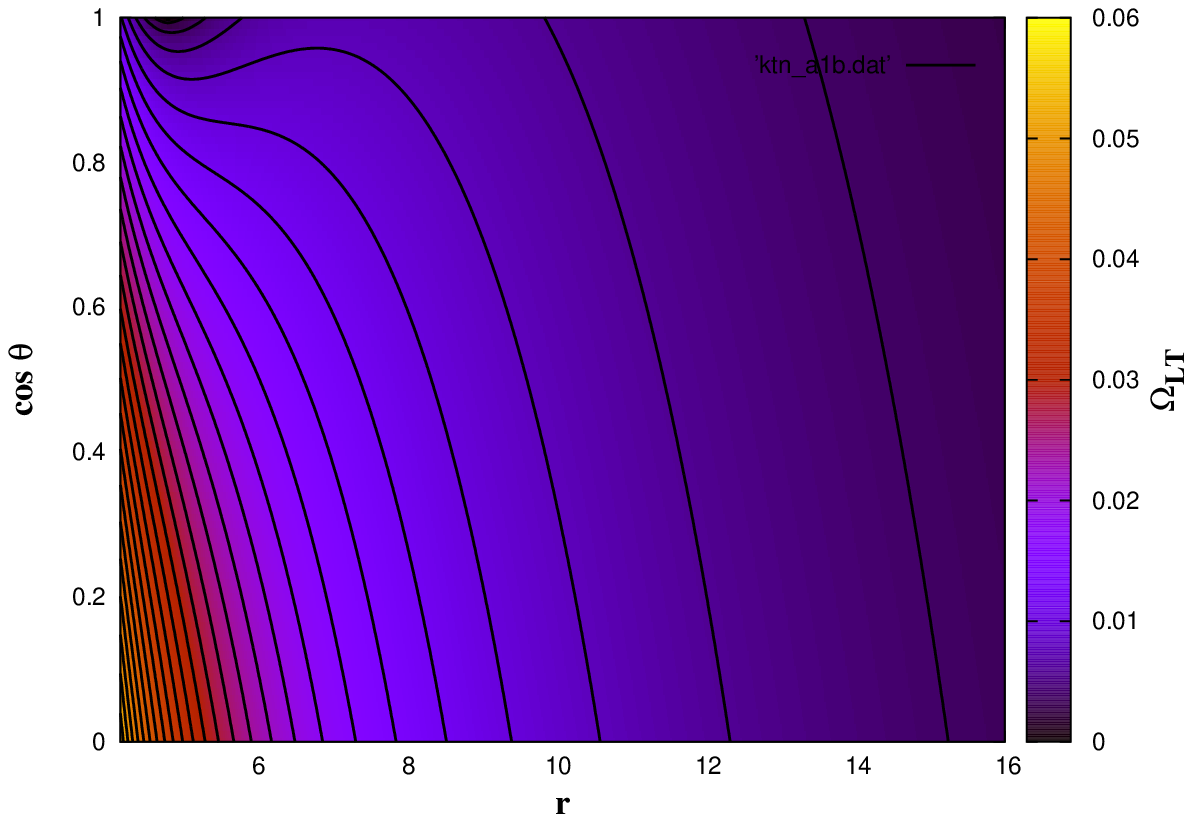}} 
\caption{\label{an1}\textit{Plot of $\O_{LT}$ vs $r$
and $\cos\th$ in the KTN spacetime for $a=n=M=1$ \cite{cc2}}}
\end{center}
\end{figure} 
In contrast to the Kerr spacetime, where the 
source of the LT precession is the Kerr
parameter $a$, the Kerr-Taub-NUT spacetime 
has an extra somewhat surprising feature :
the LT precession does not vanish even for 
vanishing Kerr parameter $a=0$, so long as the NUT charge $n \neq 0$. 
This means that though the orbital angular 
momentum $(J)$ of this spacetime vanishes, 
the spacetime does indeed exhibit an  intrinsic {\it spinlike} 
angular momentum (at the classical level itself). One can show that 
inertial frames are dragged along this orbitally
 {\it non-rotating} NUT spacetime with the precession rate \cite{cm}
\begin{eqnarray}
 \vec{\O}^{MTN}_{LT}=\f{n(r^2-2Mr-n^2)^{\f{1}{2}}}{(r^2+n^2)^{\f{3}{2}}}\hat{r}.
\label{tn1}
\end{eqnarray}

\begin{figure}[h]
    \begin{center}
\includegraphics[width=1.9in]{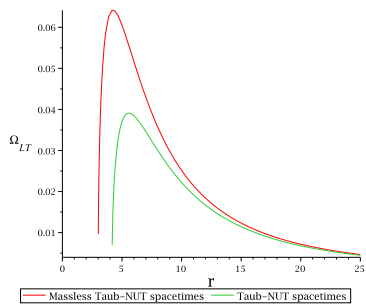}
      \caption{Plot of $r$ vs $\O_{LT}$ for $n=3$ \& $M=1$}
      \end{center}
~\label{fig}
\end{figure}

\subsection{Exact LT precession rate inside the rotating neutron stars}

The rotating equilibrium models considered in this 
paper are stationary and axisymmetric. Thus we can write the metric
inside the rotating neutron star as the following Komatsu-Eriguchi-Hachisu
(KEH)~\cite{keh} form:
\begin{equation}
 ds^2=-e^{\g+\s}dt^2+e^{2\a}(dr^2+r^2d\th^2)+e^{\g-\s}r^2\sin^2\th(d\phi-\o dt)^2
\label{met}
\end{equation}
where $\g,\,\s,\,\a,\,\o$ are the functions of $r$ and $\th$ only.
In the whole paper we have used the geometrized unit ($G=c=1$).
We assume that the matter source is a perfect fluid with a
stress-energy tensor given by
\begin{equation}
 T^{\mu\nu}=(\r_0+\r_i+P)u^\mu u^\nu+Pg^{\mu\nu}
\end{equation}
where $\r_0$ is the rest energy density, $\r_i$ is the
internal energy density, $P$ is the pressure and $u^\mu$
is the matter four velocity. We are further assuming that
there is no meridional circulation of the matter so that
the four-velocity $u^\mu$ is simply a linear combination
of time and angular Killing vectors.

In orthonormal coordinate basis, the modulus of the exact LT
precession rate inside the rotating neutron star is:
\begin{eqnarray}\nonumber
&&\O_{LT}=|\vec{\O}_{LT}(r,\th)|=\f{e^{-(\a+\s)}}{2(\o^2r^2\sin^2\th-e^{2\s})}.
\label{ltn}
\\  \nonumber           
&&\left[\sin^2\th[r^3\o^2\o_{,r}\sin^2\th+e^{2\s}(2\o+r\o,_r-2\o r \s,_r)]^2
+[r^2\o^2\o_{,\th}\sin^3\th+e^{2\s}(2\o \cos\th+\o_{,\th}\sin\th
-2\o\s_{,\th}\sin\th)]^2\right]^{\f{1}{2}}.
\\
\end{eqnarray}
The expressions of $\s (r,\th), \a(r,\th), \o(r,\th)$ have been extracted from \cite{keh}
and the frame-dragging rates have been calculated using the 
rotating neutron star {\tt rns} code \cite{str}
and plotted from the origin to the surface for some pulsars \cite{cmb}.

\begin{figure}[h!]
\begin{center}
\subfigure[along the equator]{
\includegraphics[width=1.7in,angle=-90]{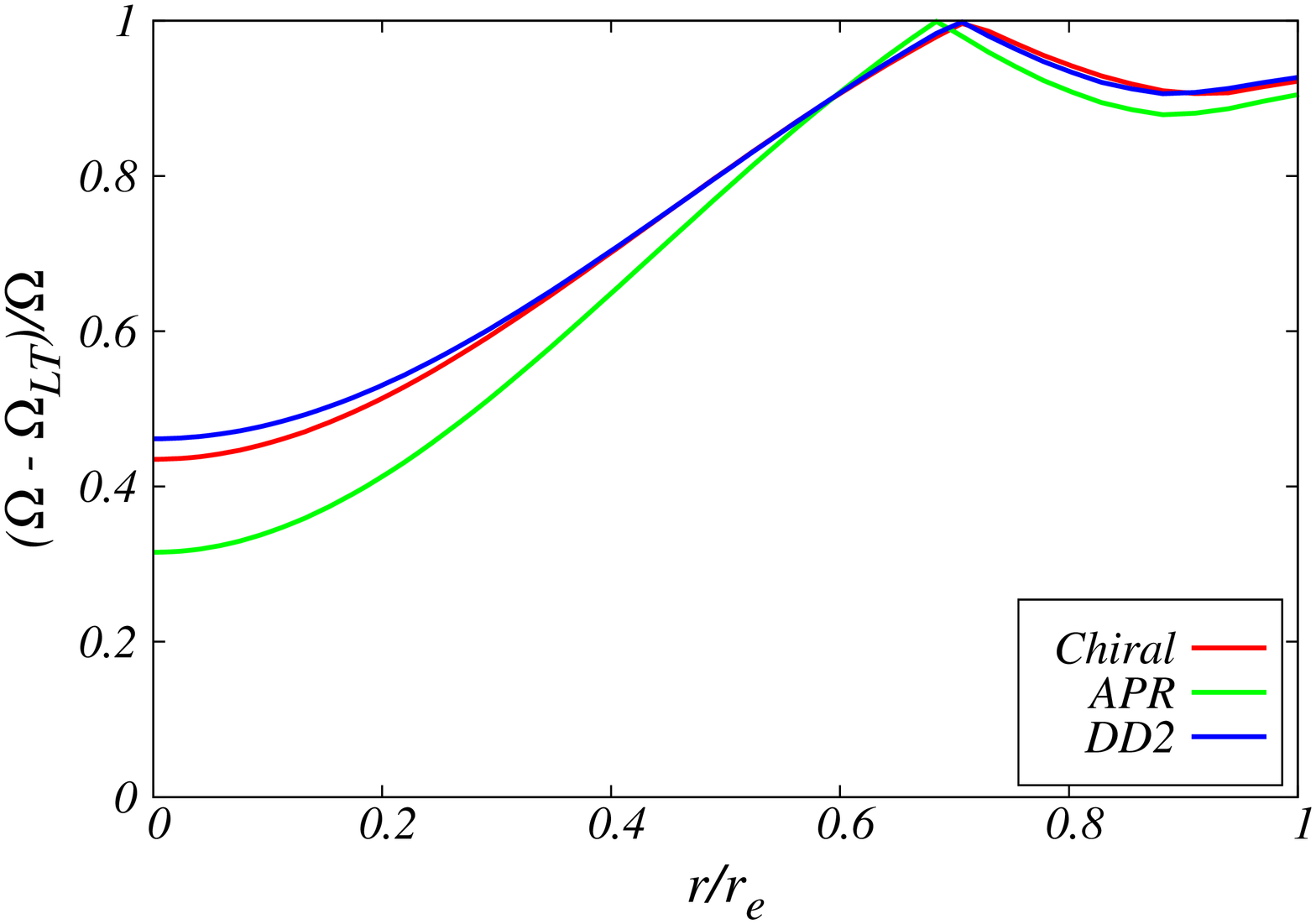}
}
\subfigure[along the pole]{
\includegraphics[width=1.7in,angle=-90]{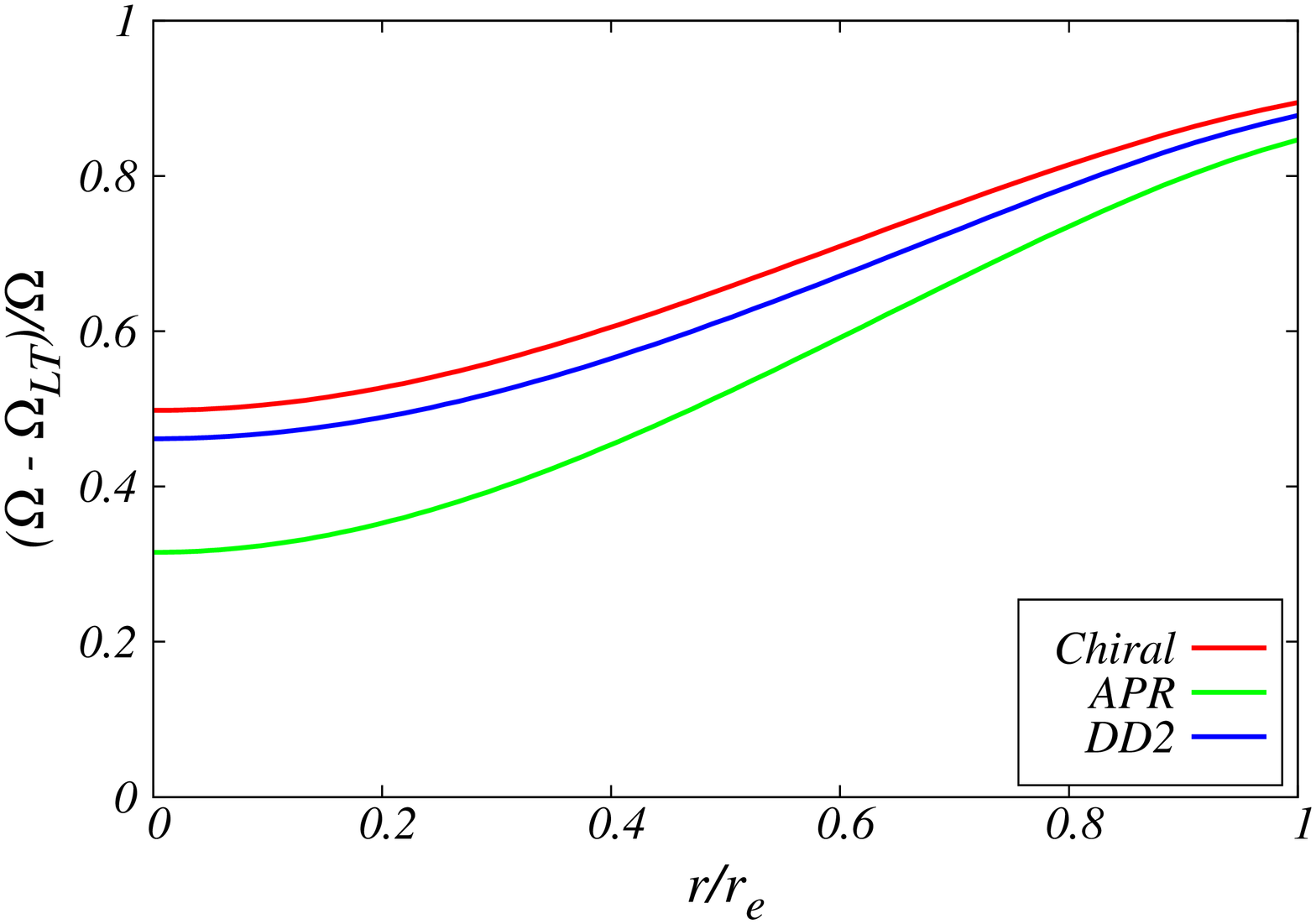}
}
\subfigure[3-D plot for APR EoS \cite{apr}]{
\includegraphics[width=1.7in,angle=-90]{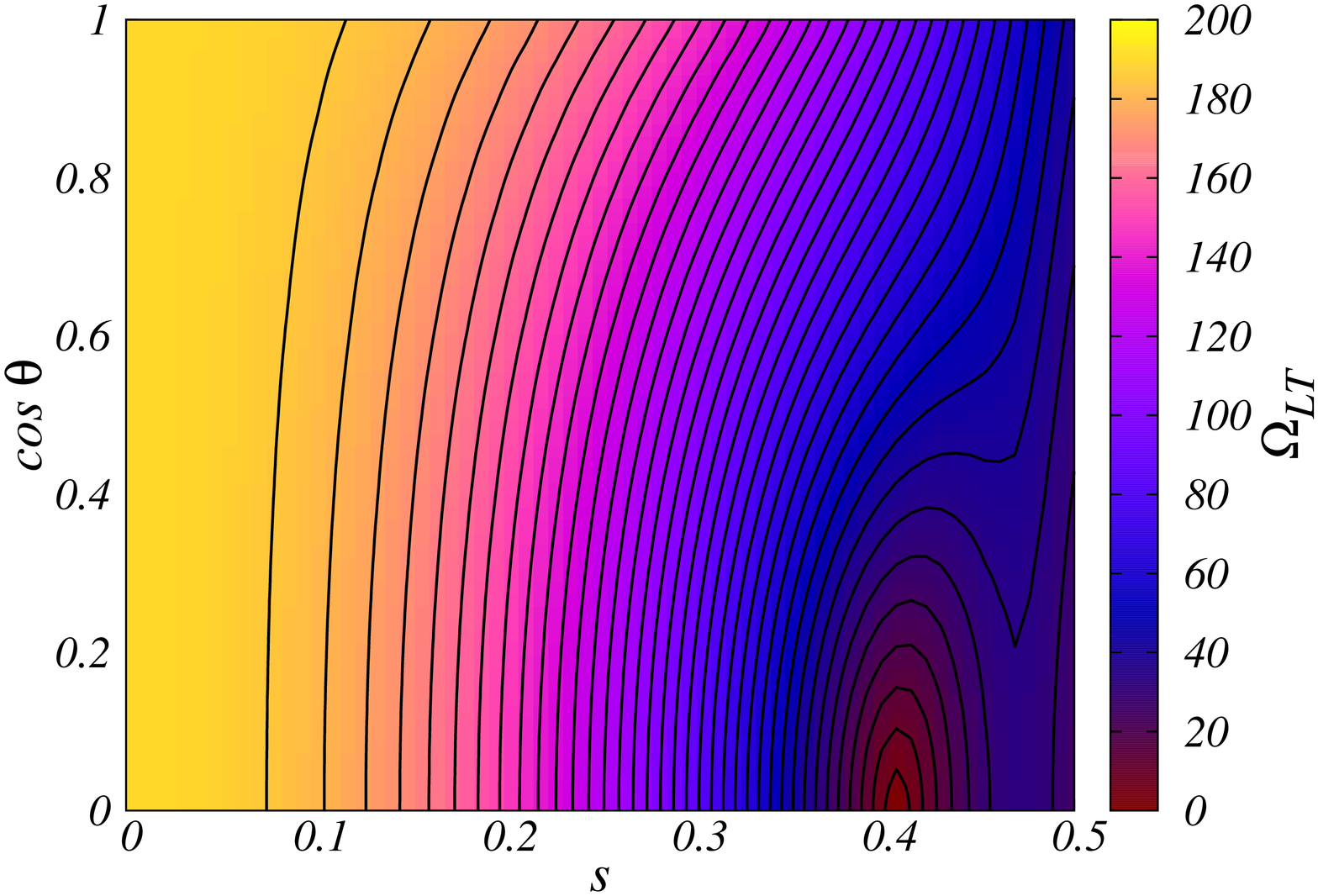}
}
\subfigure[3-D plot for $DD$2 EoS \cite{typ}]{
\includegraphics[width=1.7in,angle=-90]{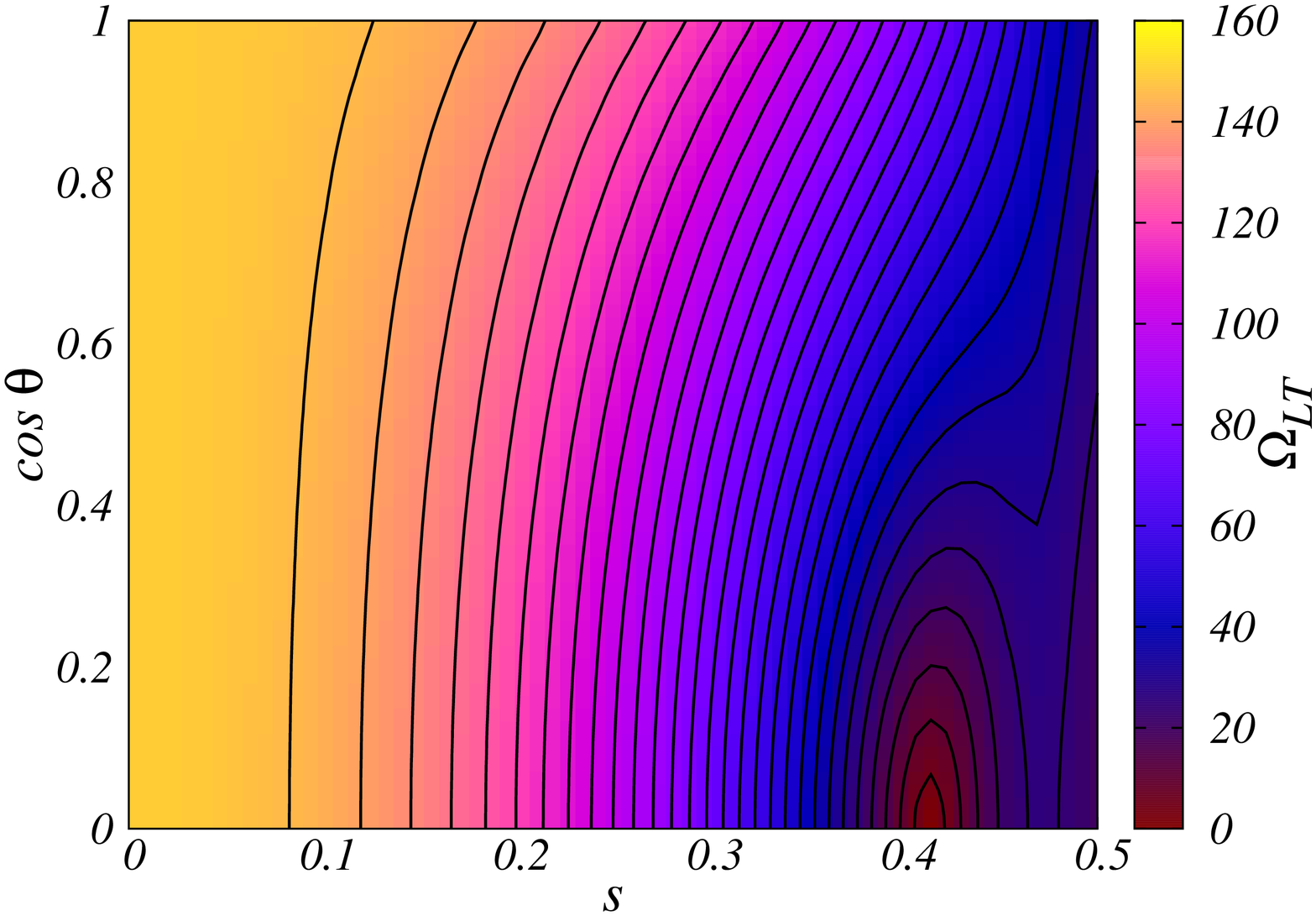}} 
\caption{\label{j0737} \textit{LT effect is calculated from the origin to the surface
in the interior of the pulsar  J0737-3039A ($M=1.337 M_{\odot}, 
\Omega = 276.8\, s ^{-1}$) (taken from \cite{cmb})}}
\end{center}
\end{figure}

\section{Conclusions \& Outlook}

In contrast to most calculations of the 
LT precession rate in the literature, which focus on the 
weak-field approximation, this article has discussed in some detail the problem of 
the exact LT precession formula in any stationary
spacetime and derived the exact LT precession formula 
for full Kerr, KTN and other spacetimes. The weak-field approximation of LT precession (Eq. 1)
for the Kerr spacetime has been then shown to emerge 
straightforwardly from our general formulation. 
Interestingly, we have shown that in the {\it non-rotating}
 and {\it spherically symmetric} Taub-NUT spacetime 
the LT precession does not vanish. 
Applying the general LT formula we have also obtained 
the exact frame-dragging rate inside the rotating neutron
stars. It is known to us that Lynden-Bell and Nouri-Zonoz \cite{lnbl}
first highlighted about the observational possibilities for NUT charges
or (gravito)magnetic monopoles and they claimed
that the signatures of such spacetime might be
found in the spectra of supernovae, quasars, or active 
galactic nuclei. In our case, the resemblance of the anomalous LT precessions 
in the KTN spacetimes and the spacetime of the pulsars could be 
the indirect proof of the existence of 
the NUT charge ((gravito)magnetic monopoles) inside the pulsars and
we also suggest that such a signature could be used to identify a role of Taub-NUT solutions
in the astrophysical observations.

Though any strong gravity measurement has not been performed till now
the analogue models of black holes can offer an alternative option of the indirect 
measurement of strong gravity LT effect in a comparatively accessible laboratory setup. We deduce precise 
estimate \cite{cgm} of the angular velocity of the precession of a test spin outside the ergoregion 
of a fluid mechanical rotating “dumb hole” in acoustic spacetimes. It is our hope that
with present technological expertise in manipulating analogue black holes, experimentalists 
will be able to successfully verify our estimate and hence, more importantly, the predicted 
strong gravity LT effect.
Accretion disk theory and the related astrophysical 
phenomena have been investigated 
using Newtonian or post-Newtonian Gravity.
The strong gravity LT effect has 
not been considered in those calculations.
As the accretion disks extend 
to the vicinity of black hole horizons associated with the very 
strong gravity regime, the effect of the frame-dragging may 
have to be taken into account. It will be interesting to
 investigate the changes to the standard accretion disc 
theory that this inclusion might entail.
The exact frame-dragging rates have already been derived 
inside and outside of the rotating neutron stars by us.
Now, it would be worth to study the quasi-periodic-oscillations 
(QPOs) in the case of accreting pulsars.
\\

{\bf Acknowledgements:}
I am highly thankful to the Scientific Organising Committee of 28th Texas
Symposium on Relativistic Astrophysics for granting me the full financial support
to attend this meeting.

%%%%%%%%%%%%%%%%%%%%%%%%%%%%%%%%%%%%%%%%%%%%%%%%%%%%%%%%%%%%%%%%%%%%%%%%%%%%
\end{document}